\documentclass[showpacs,floatfix,aps,prl,reprint,superscriptaddress]{revtex4-1}

\usepackage{amsfonts} %maths
\usepackage{amsmath}
\usepackage{amssymb}
\usepackage{braket}
\usepackage{breqn}
\usepackage{bm} 

\usepackage{dcolumn}
\usepackage{etoolbox}

\usepackage{float}
\usepackage{graphicx}
\usepackage{lipsum}
\usepackage{mathtools}

\usepackage{verbatim}
\usepackage[utf8]{inputenc} %useful to type directly diacritic characters
\usepackage{soul,xcolor,colortbl}
\usepackage{hyperref}\hypersetup{colorlinks=true,citecolor=blue,linkcolor=blue,urlcolor=blue}

\begin{document}

\title{Nucleation of Grain Boundary Phases}
\author{I. S. Winter}
\email{winter24@llnl.gov}
\author{R. E. Rudd}
\author{T. Oppelstrup}
\author{T. Frolov}
\email{frolov2@llnl.gov}
\affiliation{Lawrence Livermore National Laboratory$,$ Livermore$,$ CA$,$ 94550$,$ USA}
\date{\today}
\begin{abstract}
We derive a theory that describes homogeneous nucleation of grain boundary (GB) phases. Our analysis takes account of the energy resulting from the GB phase junction, the line defect separating two different GB structures, which is necessarily a dislocation as well as an elastic line force due to the jump in GB stresses. The theory provides analytic forms for the elastic interactions and the core energy of the GB phase junction that, along with the change in GB energy, determine the nucleation barrier. We apply the resulting nucleation model to simulations of GB phase transformations in tungsten. Our theory explains why under certain conditions GBs cannot spontaneously change their structure even to a lower energy state. 
\end{abstract}
\maketitle

The nucleation of a different phase of matter from a parent phase is one of the most basic phenomena studied within the physical sciences, and has tremendous impact on a range of technologies from the efficiency of steam engines \cite{kalikmanov2013classical} to engineering alloys \cite{khachaturyan2013theory}. Classical nucleation theory  describes the energetics of homogeneous nucleation of a solid phase in a liquid phase using only two parameters: the difference in free energy between the two phases and the interfacial energy of the two phases \cite{PorterAndEasterling}. The description of solid-solid phase transformations becomes more complicated due to the presence of elastic contributions to the nucleation energy \cite{khachaturyan2013theory}, such as the elastic effects from an Eshelby inclusion \cite{EshelbyI,EshelbyII,khachaturyan2013theory} and dislocation strain fields\cite{CAHN1957169}.

Similarly to bulk materials, interfaces can also exhibit phase-like behavior \cite{CANTWELL20141,krause2019review,cantwell2020grain}.  For interfacial phases in fluid systems, the conditions of equilibrium and stability were first derived by Gibbs \cite{gibbs1948collected}. In solid systems, phase transformations and grain boundaries (GBs), interfaces formed by two misoriented crystals of the same material, has recently become a topic of increased interest due to the accumulating experimental and modeling evidence of first-order transitions at such interfaces \cite{PhysRevLett.59.2887,PhillpotAndRickman,Rabkin1999,PhysRevB.73.024102,PhysRevLett.96.055505,DILLON20076208,Harmer182,PhysRevLett.106.046101,Divinski13,Frolov2013,RICKMAN201388,PhysRevB.92.020103,Meiners}. GB phase transitions are important because they result in discontinuous changes in grain-boundary properties such as mobility, diffusivity, and cohesive strength. \cite{DILLON20076208,wei2021direct,mishin2021interface,PhysRevLett.110.255502,Hoagland4,Luo5,Hu7}. These changes in turn can have a tremendous impact on macroscopic properties of materials such as creep and ductility by affecting a material’s microstructure \cite{DILLON2016324,Luo1999,bojarski2012changes,FRAZIER2015390,ROHRER2016231,RUPERT2016257,Khalajhedayati2016}.% \cite{RUPERT2016257,Khalajhedayati2016,DILLON20076208}.%, or influencing intergranular fracture toughness \cite{Luo5}.  
 
%Phases at interfaces, in particular, are also of great scientific and engineering interest \cite{krause2019review}.  Phase transformations at interfaces were first predicted to exist by Gibbs\cite{gibbs1948collected}.  This field has recently received renewed attention using a variety of approaches\cite{krause2019review,Meiners,Frolov2013,CANTWELL20141,Harmer182,PhysRevB.73.024102,PhysRevLett.97.075502,RICKMAN2016225}  including the recent direct experimental observation of a grain boundary phase transformation in Cu \cite{Meiners}. Grain boundary phase behavior is important to understand as shown that different grain boundary phases affect the boundary's cohesive strength, mobility, diffusivity, and sliding resistance, among other properties \cite{Hoagland4,Luo5,CANTWELL20141,Hu7,Divinski13,DILLON20075247}. As a result grain boundary phases can have a tremendous impact on properties such as creep and ductility by affecting a material's microstructure \cite{RUPERT2016257,Khalajhedayati2016,DILLON20076208} or influencing intergranular fracture toughness \cite{Luo5}. 

 While GB phase transformations  have now been seen in many different materials \cite{cantwell2020grain},  the  thermodynamics  and kinetics of these transformations  is not understood.   In principle, a grain boundary  with multiple  possible structures  of nearly the same energy should be able to sample its different states  at finite temperature,  making the  GB  structure an ensemble  of different configurations \cite{Hoagland4,HAN2016259}. However, in experimental and modeling studies GBs  behave more like conventional 3D phases: only one GB phase is observed at a time or, when a transformation  occurs, the two GB phases  are separated by  a sharp 1D interface \cite{Meiners,Frolov2013,PhysRevLett.59.2887,FrolovAndMishin,o2018grain}. Moreover, GB phase transitions can be  sluggish or delayed  even when they are not limited by solute diffusion \cite{Meiners,CANTWELL201678,SCHUMACHER2016316}.  Time-Temperature-Transformation (TTT)  GB diagrams  describing these kinetics have been proposed  as a new strategy for GB  engineering  and optimization of materials properties \cite{CANTWELL201678,SCHUMACHER2016316}. 

No nucleation theory currently exists explaining GB phase transformation behavior. A major gap is the poor understanding of the role of GB phase junctions, the 1D interfaces separating two different phases within a GB \cite{FrolovBurgers,Pond77a,Pond77b,HIRTH19964749,HIRTH2013749}. In this work, we  use the  classical nucleation theory (CNT) approach to describe GB phase transformations. The resulting theory  recognizes that GB  phase junctions  are  dislocations  as well as  elastic line forces  and  for the first time quantifies the  contribution of their elastic interactions and core energy to the nucleation barrier. 

\begin{figure}[ht!]
\centering
\includegraphics[width=\columnwidth]{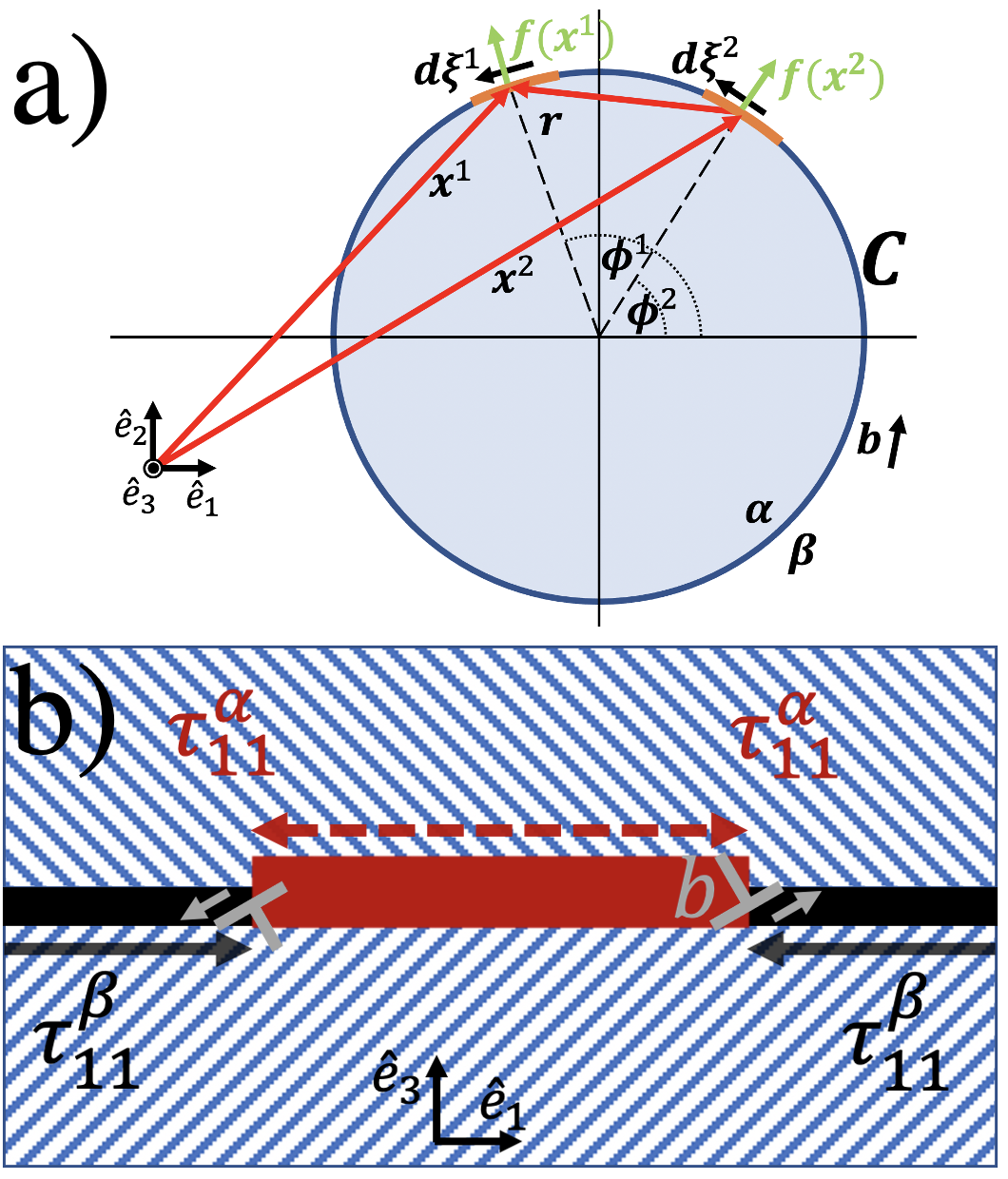}
\caption{Panel a  depicts a top view of the GB  plane showing  a  circular  nucleus  of  GB phase $\alpha$  inside the parent GB phase  $\beta$.  The contour $\bm{C}$   represents the GB  phase junction. Panel b  shows a side view  of a slice  through  the nucleus of GB phase $\beta$  showing  the misoriented bulk crystals  and GB phases.  Two GB phase junctions  are indicated  by dislocation symbols.  The imbalance of GB stresses $\bm{\tau}^{\beta}$ and $\bm{\tau}^{\alpha}$ at GB  phase junctions  also  produces line forces.}
\label{fig:illustrate_3D}
\end{figure}

Consider nucleation of a new GB phase $\alpha$  from a parent GB  phase $\beta$,  as shown in Fig.\ \ref{fig:illustrate_3D}a.    For simplicity, we assume that the nucleus is circular.  Figure \ref{fig:illustrate_3D}b  shows the side  view of the system in \ref{fig:illustrate_3D}a  which also includes the two  misoriented  crystals. GB  phase junctions  are dislocations  as well as line forces, arising from the imbalance of the grain boundary stresses $\bm{\tau}^{\beta}$ and $\bm{\tau}^{\alpha}$: $\bm{f}=(\bm{\tau}^{\alpha}-\bm{\tau}^{\beta})\cdot \hat{\bm{r}}$, where $\hat{\bm{r}}$ is the radial unit vector. The GB phase junction  indicated on the figure as a contour $\bm{C}$  forms a closed  circular loop, which is a dislocation loop of Burgers vector $\bm{b}$ \cite{FrolovBurgers} and line force $\bm{f}$. The energy of the nucleus is  given by
\begin{equation}\label{eq:nuc_3D_gb}
    E^{nuc}(R) = \pi R^2 \Delta \gamma^{\alpha \beta} + 2\pi R \bar{\Gamma}^{\alpha \beta} + E^{d d}(R) + E^{d p}(R) + E^{p p}(R),
\end{equation}
 Here the first two terms are the usual CNT contributions describing the driving force  for the transformation due to the reduction in the  GB free energy  per unit area $\Delta \gamma^{\alpha\beta}=\gamma^{\alpha}-\gamma^{\beta}$ and  the  increase due to the  perimeter energy,  $\bar{\Gamma}^{\alpha \beta}$,  the effective orientation-averaged core energy per unit length of the GB  phase junction.  The last three terms  on the right hand side of Eq.\ (\ref{eq:nuc_3D_gb})  describe the elastic part of the GB phase junction energy. $E^{dd}(l)$ is the elastic self-energy of the dislocation loop, $E^{pp}(l)$ is the elastic energy of the line force loop, and $E^{dp}(l)$ is the interaction between the dislocation and line force loop.  A detailed derivation is given in the Supplemental Materials. Each elastic term is given as:

\begin{subequations}\label{eq:3DelasticTerms}
\begin{dmath}
E^{dd}(R) = \frac{\mu R}{4(1-\nu)} \left[  \Big(2b_3^2 +  (b^p)^2(2-\nu)\Big)\ln\left(\frac{4R}{\rho}\right) - 2\Big(b_3^2 + (b^p)^2(2-\nu)\Big)\right],
\end{dmath}

\begin{dmath}
E^{dp}(R) = -\frac{(f_1+f_2)b_3R}{2(1-\nu)}\Big[(1-2\nu)\ln (4R/\rho) -3 + 4\nu \Big],
\end{dmath}

\begin{dmath}
E^{pp}(R) = -\frac{R}{32\mu(1-\nu)}\Big[\left(f^2(13-16\nu)-2f_1f_2\right)\ln (4R/\rho)-4f^2(7-8\nu)\Big].
\end{dmath}
\end{subequations}

\noindent $(b^p)^2 = b_1^2+b_2^2$. $\rho$ is the core radius of the dislocation, and is $b$. $b_3$ is the $\hat{\bm{e}}_3$ component of the Burgers vector; and  $f^2 = f_{1}^2 + f_{2}^2$, $f_{1} = (\tau_{11}^{\alpha}-\tau_{11}^{\beta})$, and  $f_{2} = (\tau_{22}^{\alpha}-\tau_{22}^{\beta})$.

Equations (\ref{eq:nuc_3D_gb}) and (\ref{eq:3DelasticTerms})  describe the energetics of GB phase  nucleation  by predicting the energy of the nucleus as a function of its size.  Equation (\ref{eq:nuc_3D_gb}) incorporates  the elastic energy  due to the GB phase junctions and can be used  to predict the size of the critical  nucleus.  In the  3D   system considered here, the perimeter of the GB phase junction  changes with the radius of the nucleus;  as a result, the core  energy $\bar{\Gamma}^{\alpha\beta}$   cannot be calculated directly  from MD using this 3D model, but could be treated as a fitting perimeter.  Before we proceed to the MD part of this study, we  show that $\bar{\Gamma}^{\alpha\beta}$  can be calculated  directly from molecular statics simulations of quasi-2D  nucleation.

Consider a GB phase transformation in a quasi-2D  system  such as a thin film, schematically shown in Fig.\ \ref{fig:illustrate_3D}b.  Here the  length of each  GB phase junction is fixed  and set by the  film  thickness, $L$.  In this 2D case,  the two GB  phase junctions  are  dislocations and line forces,  and the energy of the system per unit thickness as a function of  nucleus  width $l$  is given by

\begin{equation}\label{eqn:nucleation}
    \frac{E }{L}^{2D}= \Delta \gamma^{\alpha \beta} l + \frac{E}{L}^{d d}(l) + \frac{E}{L}^{p p}(l) + \frac{E}{L}^{d p}(l) + 2\Gamma^{\alpha \beta},
\end{equation}

This is the 2D analog of Eq.\ (\ref{eq:nuc_3D_gb}). In Eq.\ (\ref{eqn:nucleation}) the  core  energy  $\Gamma^{\alpha\beta}$ is decoupled  from the elastic energy terms.  Since $E^{2D}$  and $\Delta \gamma^{\alpha \beta}$  can be calculated directly  from  molecular statics calculations and the  elastic energy terms can be  evaluated using the elasticity theory, Eq.\ (\ref{eqn:nucleation})  and  quasi-2d MD  simulations of  nucleation   can be used to calculate $\Gamma^{\alpha\beta}$.

%\begin{widetext}
%\begin{figure}[ht!]
%\centering
%\includegraphics[scale=0.35]{figures/grain_boundary_phase_illustration.png}
%\caption{Illustration of the important interactions to consider in the homogeneous nucleation of the $\beta$ grain boundary phase (brown region). Apart from the difference in grain boundary energy between the two phases, there are also elastic interactions that arise from the presence of dislocations (due to geometric constraints) and line forces (due to differences in grain boundary stress) at the phase boundaries.}
%\label{fig:gbphase_illustration}
%\end{figure}
%\end{widetext}

The three elastic energy terms are derived in the Supplemental Materials:
\begin{subequations}\label{eqn:elastic_energies}
\begin{align}
    \frac{E}{L}^{dd}(l) &=  \frac{\mu}{2\pi}\left( b_2^2 + \frac{(b^e)^2}{1-\nu}\right)\ln\left(\frac{l}{\rho}\right) + C^{dd}+\Gamma^{dd},\\
    \frac{E}{L}^{dp}(l) &= -\frac{f_1 b_3 (1-2\nu) }{2\pi(1-\nu)} \ln\left(\frac{l}{\rho}\right) + C^{dp}+\Gamma^{dp},\\
    \frac{E}{L}^{pp}(l) &= -\frac{f_1^2(3-4\nu)}{8\pi \mu (1-\nu)} \ln\left(\frac{l}{\rho}\right) + C^{pp}+\Gamma^{pp},
\end{align}
\end{subequations}
where $C^{dd}$, $C^{dp}$ and $C^{pp}$ are elastic terms that are not dependent on $l$ and the $\Gamma$ terms are the core energy terms associated with each interaction.

Combining the three elastic interaction terms together with the difference in grain boundary energy between the two phases, the nucleation energy is expressed as
\begin{dmath}\label{eqn:classical}
    \frac{E}{L}^{2D} = \Delta \gamma^{\alpha\beta} l + \frac{1}{8\pi(1-\nu)}\left( 4\mu \Big[(1-\nu)b_2^2 + (b^e)^2\Big] - \frac{f_1^2 (3-4\nu)}{\mu} - 4f_1b_3 (1-2\nu)\right) \ln \left(\frac{l}{\rho}\right) + 2\Gamma^{\alpha\beta}+C,
\end{dmath}
with $C$ being the elastic terms not dependent on $l$.

Equation (\ref{eqn:classical})   predicts that when the elastic interactions are included,   the quasi-2D nucleation  energy   is no longer a decreasing function of $l$  for a negative $\Delta \gamma^{\alpha\beta}$,  but it increases first  for small nuclei  resulting in a nucleation barrier.   This nucleation  barrier allows GBs to remain  in a  metastable state even in quasi-2D systems like thin films. The critical length of the $\alpha$ nucleus is determined by solving $\frac{d E^{2D}(l)}{dl} = 0$ for $l$, which gives an analytical solution: 
\begin{equation}\label{eqn:lcrit}
    l^{c} = \frac{4\Big[(1-\nu)b_2^2 + (b^e)^2\Big] \mu^2 - 4 \mu f_1 b_3 (1-2\nu) - f_1^2(3-4\nu)}{8\pi \mu\Delta \gamma^{\alpha\beta} (\nu-1)}.
\end{equation}

 The analysis presented so far shows that in both 3D and quasi-2D  cases,  the elastic interaction  energy due to  GB  phase junctions  can increase the GB transformation barrier. These nucleation barriers  stabilize metastable GB structures from transforming to their ground states. They  can also prohibit  sampling of other GB states with the same or higher energy by fluctuations at finite temperature. The energy of the elastic interactions can have a significant influence on  the transformation behavior  away from GB  critical points and  near  the  equilibrium  coexistence where  $\Delta \gamma^{\alpha\beta}$  is small.  
  To test the predictions of our theory  we  performed MD  simulations of GB phase  transformations.  Using  the derived theory  we extract the GB phase junction core energy from the MD data, calculate the elastic energy contribution to the nucleation  barrier, the critical  nucleus size and  make a direct comparison with the MD results.

 As an example we have selected  the $\Sigma29(520)[001]$ symmetric tilt grain boundary in tungsten (W) modeled with the EAM potential developed by Zhou \textit{et al.} \cite{ZHOU20014005}. W  was chosen because it is  elastically isotropic,  so  its elastic energy should be described well by the developed theory. The shear modulus and Poisson ratio were found to be $\mu = 160.0$ GPa and $\nu=0.280$ respectively. Previously, grand-canonical GB structure searches demonstrated multiple GB  phases in several different W GBs \cite{FROLOV2018123,frolov2018grain}. We selected  this particular boundary  because the GB  structure search  identified  two distinct GB  structures  that correspond to  two different  grain translations, but are composed of the same number of atoms \cite{BISHOP1968133,SuttonI,SuttonII}. The latter property is convenient  for our analysis because it allows us to create nuclei of a new GB phase at 0 K  to calculate their energy and study GB  phase  transformations on the short MD times scale.

The $\alpha$ phase is the ground  state with the 0 K energy $\gamma^{\alpha}=2.342$ J$/$m$^2$ and  the $\beta$ phase is metastable with energy $\gamma^{\beta}=2.418$ J$/$m$^2$.  To study the finite-temperature stability of both GB structures  we performed MD  simulations of  each boundary using the NVT ensemble  at temperatures of 1500K, 2000K and 2500K.  Initially  we used   relatively large GB  areas with dimensions of $30 \sqrt{29}a_0\times 10 a_0$. 

\begin{figure}[ht!]
\centering
\includegraphics[width=\columnwidth]{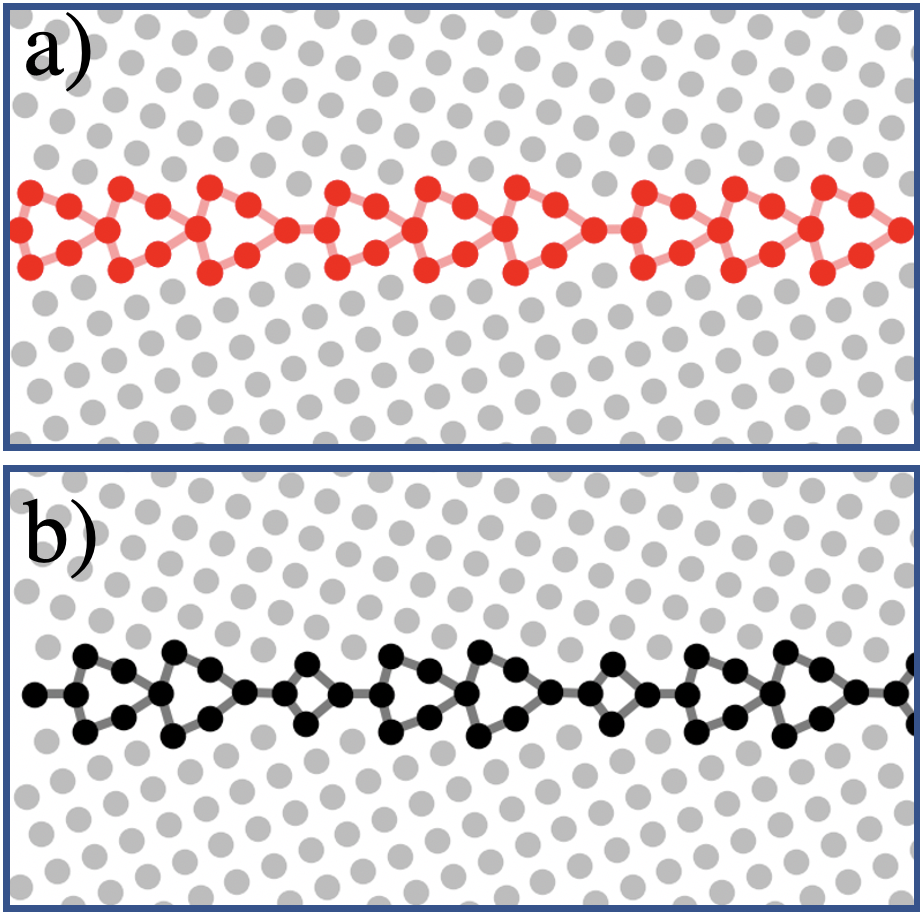}
\caption{Atomic structure of the two grain boundary phases. Panel a shows the $\alpha$ phase (ground state) and panel b shows the $\beta$ phase (metastable).}
\label{fig:example_structures}
\end{figure}

The structures of the two grain boundary phases are shown in Fig.\ \ref{fig:example_structures}. In these simulations  both GB  structures  remained stable and did not transform even after 120 ns of simulation at the  highest  temperature of 2500K.   This already suggests  that the transformation barriers  are significant  compared to the driving forces for nucleation and growth, such as those associated with GB  free energy difference or the thermal  fluctuations.

To observe the GB phase transformations  in MD  we reduced the $\hat{\bm{e}}_2$ dimension of the simulation block  to the smallest possible value equal to just one lattice parameter making it quasi-2D. In this case  the nucleation barrier is reduced significantly and after only 8 ns of MD simulation at 1500K  the $\beta$  phase  transformed  into the $\alpha$ phase  by  nucleation and growth.  The same behavior was observed at higher temperatures,  confirming that the  $\beta$  phase  is metastable  within the entire temperature range. These  simulations clearly  demonstrate that the  GB  structure  observed in MD and GB phase  transformation behavior is very sensitive to the choice of GB area,  because for small dimensions  periodic boundary conditions  artificially influence the  nucleation barrier. 
We used the simulation block  containing both phases  to calculate the  Burgers vector of the  GB phase junction following the methodology  described in Ref.\ \cite{FrolovBurgers}.   The Burgers vector  was found to be $\bm{b} = 0.600\textrm{\AA} \hat{\bm{e}}_1 + 0.423\textrm{\AA}\hat{\bm{e}}_3$, which has the same order of magnitude as a disconnection for this grain boundary: $\bm{b}^{DSC} = -0.416 \textrm{\AA} \hat{\bm{e}}_1 - 0.416\textrm{\AA}\hat{\bm{e}}_3 $ \cite{GRIMMER19741221,HAN2018386}. The tangential component of $\bm{b}$ is due to the different grain  translations (or excess GB shears\cite{PhysRevB.85.224106}) of the two GB phases and  the normal component  is equal to the  difference in GB excess  volumes \cite{FrolovBurgers,Pond77a,Pond77b}. The details of the  Burger circuit analysis  are included in the Supplemental Materials.

 To explain the surprising stability  of the  metastable GB phase $\beta$  observed in the full 3D MD simulations,  we could use our theory and predict the  nucleation barriers. This would require  calculations of finite-temperature elastic  constants  and free energies  of all the defects involved including the free energy of GB junction cores. Some of those calculations are nontrivial and are beyond the scope of this study.
 However, we can still get valuable insights  into the energetics  of the GB  phase  transformation  and  validate our nucleation  model  by performing molecular statics calculations at 0 K.  Knowing the  Burgers vector  of the  GB  phase junction from the MD analysis, GB energies,  GB stresses  and the material's elastic constants  allows us to predict the nucleus energy  as a function of its  size. The last missing  ingredient  of the  nucleation  theory is the core energy of the GB  phase junction $\Gamma^{\alpha\beta}$.  
 To obtain $\Gamma^{\alpha\beta}$ we start our  molecular statics  analysis from the  quasi-2D geometry.  The simulation block with the GB  had dimensions $120 \sqrt{29} a_0\times a_0\times 10 \sqrt{29} a_0$. With the small dimension along the GB tilt axis as before. By  applying the appropriate  initial translations  to the upper and  lower   grains  followed by  an energy minimization  we prepared  bicrystals  with the  parent phase $\beta$  containing  a  nucleus of phase $\alpha$  of different sizes.  The method for constructing the grain boundary nucleus of size $l$ is described in the Supplemental Materials. The  nucleation energy  is then calculated  as the difference  between total  energies  of the system with  and  without the nucleus. 
 
\begin{figure}[ht!]
\centering
\includegraphics[width=\columnwidth]{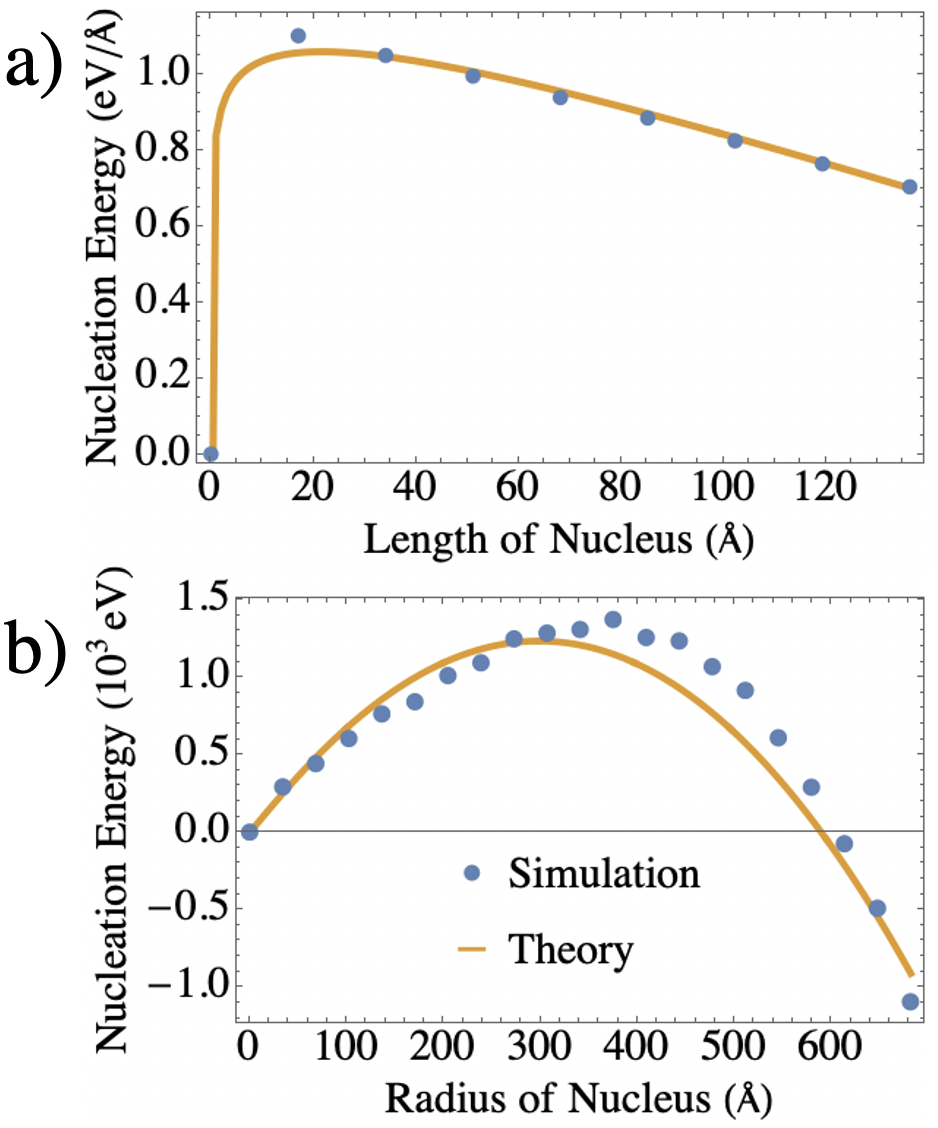}
\caption{Energy of the GB phase, $\alpha$, nucleus created inside metastable GB phase, $\beta$, as a function of its size for a) quasi-2D and b) fully-3D geometries at 0K. The blue points correspond to the direct calculations using molecular statics, while the nucleation energy predicted by the model developed in this work is shown as a solid line. }
\label{fig:nucleation_molecular_statics}
\end{figure}

 Figure \ref{fig:nucleation_molecular_statics}a  shows an excellent agreement between the  nucleation  energy  calculated directly from molecular statics (discrete blue points) and the predictions of the  nucleation model (solid line)  developed in this work. The critical nucleus  for $\Sigma29(520)[001]$ from the molecular statics simulations was found to be $l^{MD}$=17.0 \AA. $l^c=21.6$ \ \AA\ is  predicted from Eq.\ (\ref{eqn:lcrit}).  More importantly,  Eq.\ (\ref{eqn:classical}) and the MD   data predict $\Gamma^{\alpha\beta}_{min} =  0.383$\ eV$/$\AA\  for this geometry  when the  GB phase junction is  parallel to the tilt axis.  Similar to regular  dislocations, the core energy   $\Gamma^{\alpha\beta}$  may strongly  depend on the line direction.  By performing several quasi-2d calculations  changing the GB  phase junction direction, we found that $\Gamma^{\alpha\beta}_{min} =  0.383$\ eV$/$\AA\ and $\Gamma^{\alpha\beta}_{max} = 1.73$\ eV$/$\AA, where $\Gamma^{\alpha\beta}_{min}$ corresponds to $\bm{\xi} = [001]$ and $\Gamma^{\alpha\beta}_{max}$ corresponds to $\bm{\xi} = [\bar{2}50]$. To the best of our knowledge,  these are the first  reported  values of GB  phase junction core energy, and its directional anisotropy.

\begin{table}
\caption{\label{tab:facet_210_310} The relevant grain boundary properties for the two W GB phases. The other material properties are $\mu = 160.0$ GPa, $\nu=0.280$ and $\Gamma ^{\alpha \beta}$ = 0.383 - 1.73 eV/\AA.}
\begin{ruledtabular}
\begin{tabular}{cccc}
grain boundary & $\gamma$ (J/m$^2$)  & $\tau_{11}$  (J/m$^2$) & $\tau_{22}$ (J/m$^2$) \\
\hline
$\alpha$  & 2.342 & 3.926 & 3.609\\
$\beta$ & 2.418 & 0.138 & 5.775
\end{tabular}
\end{ruledtabular}
\end{table}

%\begin{figure}[ht!]
%\centering
%\includegraphics[scale=0.4]{figures/s29_520_001.png}
%\caption{Comparison of molecular dynamics simulation and classical nucleation theory plots of the dependence of the nucleation energy on length of the nucleus for a $\Sigma 29(520)[001]$ symmetric tilt grain boundary.}
%\label{fig:nucleation_molecular_statics}
%\end{figure}

Next, we tested the accuracy of the 3D model,  by comparing its predictions to   results of molecular statics  simulations of  circular  nuclei. The computational details of these simulations are  given  in  the  Supplemental  Materials. In this case, the GB had a roughly square shape  with the block dimensions. The system contained 8.7 million atoms.
 The results of the analysis are  shown in  Fig.\ \ref{fig:nucleation_molecular_statics}b. The nucleation energy predicted by Eq.\ (\ref{eq:nuc_3D_gb}) compares extremely well with the simulation results. For instance, the critical radius of the nucleus is predicted to be $R^c = 299$ \AA, from the MD data the critical nucleus is found to have a radius of approximately 375 \AA. The large nucleation barrier and critical nucleus size  are consistent with the observation that GB transformation did not occur in our fully 3D simulations even at higher temperatures.
 
The critical nucleus sizes of the quasi-2D and fully 3D cases are vastly different with $l^c = 21.6$\ \AA\ (2D) compared to the critical diameter $D^c = 598$\ \AA (3D).  This difference is not surprising because the contributions  to the nucleation barrier  scale differently with the nucleus size  for these two geometries. However, this stark difference in critical radius points to the importance of simulating fully three dimensional systems for grain boundary phase nucleation. The consequences are borne out by MD simulation. In a thinner quasi-2D system, the transformation behavior depends strongly on the GB phase junction length. The $\alpha$ phase will nucleate at 1500~K after 8~ns with a repeat length of $a_0$ along $\hat{\bm{e}}_2$, but no nucleation is seen even after 120 ns, when the repeat length is increased to $2 a_0$. 
 
%\begin{widetext}
\begin{figure}[ht!]
\centering
\includegraphics[width=\columnwidth]{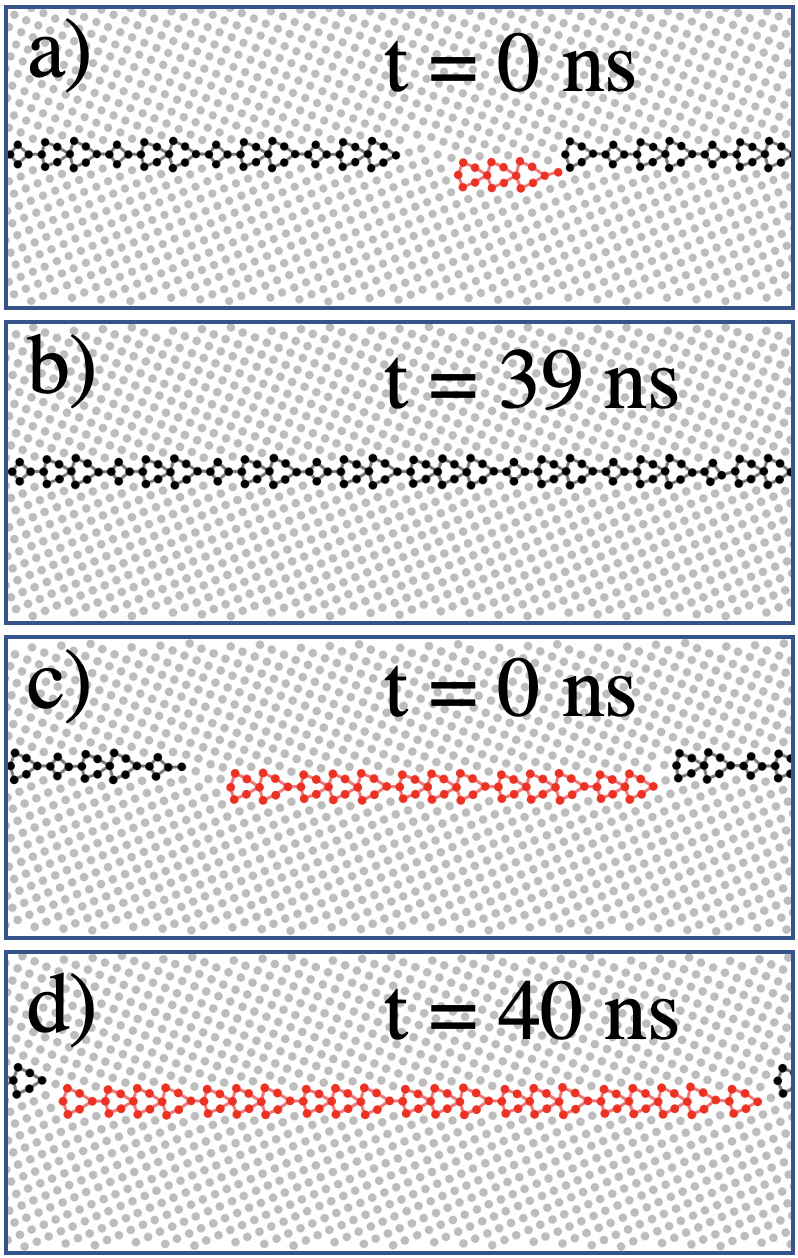}
\caption{The evolution of a dual-phase grain boundary system in MD simulation at T=1500K in a quasi-2D geometry. Panels a and b depict the shrinkage and ultimate annihilation of a subcritical length nucleus. Panels c and d depict the growth of a supercritical length nucleus. These simulations confirm the presence of nucleation barrier due to elastic interactions.}
\label{fig:shrinkage_and_growth}
\end{figure}
%\end{widetext}

To further show that predictions of our nucleation model are not merely applicable to 0 K, we also performed high-temperature MD simulations. Specifically, using the quasi-2D geometry we prepared a dual phase system with a  nucleus of the  lowest  free energy phase $\alpha$  embedded  inside the   metastable  phase $\beta$.  In one system the length of the $\alpha$ phase was less than the critical nucleation length and in the other it was greater than the critical length  estimated at 0 K; cf.\ Figure \ref{fig:shrinkage_and_growth}a and \ref{fig:shrinkage_and_growth}b, respectively. We then allowed the systems to evolve at 1500 K. As shown from Fig.\ \ref{fig:shrinkage_and_growth}a, the $\alpha$ nucleus of subcritical length shrinks and ultimately disappears from the simulation over the course of 40 ns. On the other hand, the nucleus of supercritical size grew, transforming the metastable boundary into its ground state $\alpha$, as shown in  Fig.\ \ref{fig:shrinkage_and_growth}b. These simulations confirm the presence of a nucleation barrier due to elastic interactions and show that the GB phase $\beta$ can remain metastable even in a quasi-2D geometry.

Here we have derived a theory that describes homogeneous nucleation of GB phases. Our analysis recognizes that any GB phase junction necessarily contains a dislocation and a line force resulting from an imbalance in GB stresses. The theory quantifies the contributions from the elastic interactions and the core energy of the GB phase junction to the nucleation barrier. The predictions of the theory are in excellent agreement with the direct MD calculations. We find that the nucleation barriers are significant with the elastic interaction energy contributing  about $25\%$,  with the rest of the energy coming from  the core energy of the junction.

Our finite-temperature MD simulations have shown that both GB phases studied remain stable and do not transform even at elevated temperatures, which is consistent with the large transformation barriers calculated at 0K. Our theory explains why away from critical points GBs cannot spontaneously change their structure even to a lower energy state.  By quantifying the nucleation barrier one can in principle use the model to predict limits of metastability of different GB structures. While the analysis developed here has been applied to one particular boundary in tungsten, previous studies of GB phase transformations reported long nucleation times, sharp GB phase junctions and well defined nuclei \cite{Meiners,Frolov2013,PhysRevLett.59.2887,FrolovAndMishin,o2018grain,FROLOV2018123,frolov2018grain}, which suggests that the conclusions of this study are general.
 
 %While the analysis developed here has been applied to one particular boundary in tungsten, previously published simulations\cite{Meiners,Frolov2013,FROLOV2018123,frolov2018grain} and direct experimental observations of structural transformations at GBs\cite{Meiners} suggest that the conclusions of this study are general.  Our results are consistent with prior GB studies that reported long nucleation times, sharp GB phase junctions and well defined nuclei.
 
Beyond first-order GB transformations discussed in this work, the analysis helps us better understand finite-temperature behavior of GBs in general. Prior studies suggested that at finite-temperature GBs sample higher energy states with Boltzmann probability, so that the GB structure is not unique but rather is represented by a properly weighted ensemble of different structures \cite{Hoagland4,HAN2016259}.  Our study shows that the energy difference in the Boltzmann factor should also include the energy of the GB phase junction, in addition to the energy difference per unit area. At lower temperatures this positive energy can suppress phase fluctuations, resulting in the unique GB structure often observed in experiments.

%GB disconnections,  triple junctions,  facet junctions  and surfaces  may act as  effective sites for heterogeneous nucleation, by providing a release to the elastic field of the nucleus or changing the core energy contribution.  
By recognizing  and quantifying the  elastic  energy of GB phase junctions,  our study creates a foundation upon which further heterogeneous nucleation models can be developed.

\section{Acknowledgements}
This work was performed under the auspices of the U.S. Department of
Energy by Lawrence Livermore National Laboratory under Contract DE-AC52-07NA27344. The work was funded by the Laboratory Directed Research and Development Program at LLNL under tracking number 19-ERD-026.

\bibliography{aapmsamp}% Produces the bibliography via BibTeX.

\end{document}